\magnification=\magstep1
\centerline{Single Close Encounters Do Not Make Eccentric Planetary Orbits}
\bigskip
\centerline{J. I. Katz}
\centerline{Racah Institute of Physics, Hebrew University, Jerusalem 91904,
Israel}
\centerline{and}
\centerline{Department of Physics, Washington University, St. Louis, Mo. 63130,
U. S. A.}
\centerline{katz@wuphys.wustl.edu}
\bigskip
\centerline{Abstract}
\medskip
The recent discovery of a planet in an orbit with eccentricity $e = 0.63 \pm
0.08$ around the Solar-type star 16 Cyg B, together with earlier discoveries
of other planets in orbits of significant eccentricity, raises the question of 
the origin of these orbits, so unlike the nearly circular orbits of our
Solar system.  In this paper I consider close encounters between two 
planets, each initially in a nearly circular orbit (but with sufficient 
eccentricity to permit the encounter).  Such encounters are described by a 
two-body approximation, in which the effect of the attracting star is 
neglected, and by the approximation that their separation vector follows a 
nearly parabolic path.  A single encounter cannot produce the present state 
of these systems, in which one planet is in an eccentric orbit and the other 
has apparently been lost.  Even if the requirement that the second planet be
lost is dropped, nearly circular orbits cannot scatter into eccentric ones.
\vfil
\eject
\centerline{1. Introduction}
\medskip
The recent discovery (Cochran, {\it et al.} 1996) of a planet orbiting
16 Cyg B with an eccentricity $e = 0.63 \pm 0.08$, combined with the
earlier discoveries of a planet orbiting 70 Vir with $e = 0.40$ (Marcy \& 
Butler 1996) and of one orbiting HD 114762 with $e = 0.35$ (Mazeh, Latham \&
Stefanik 1996), raise the question of how these eccentricities are 
produced.  It is conventional to assume that planets form from a disc of
gas and particles whose orbits have been circularized by frictional
(viscous) dissipation.  This is supported by the fact that the observed 
eccentricities of planets in the Solar System are generally quite small;
even their finite but small eccentricities are probably explicable as the
result of mutual gravitational perturbations.  There are two apparent
exceptions to this rule of small eccentricities in the Solar System: Pluto,
with $e = 0.250$, resembles a strayed asteroid more than a planet; its
present orbit is strongly affected by its 3:2 orbital resonance with Neptune,
and its previous history may reflect a close encounter with Neptune (and
Triton).  Mercury has $e = 0.206$, which is less obviously explicable.

One possible explanation for large planetary eccentricities is 
gravitational interaction between two planets initially in nearly circular
orbits, one of which was left in the observed eccentric orbit and the
other expelled.  Expulsion (either to infinity or into a very long period
nearly parabolic orbit) is required because the data exclude a second planet
with mass and period comparable to those of the observed planet.  Conservation
of energy and angular momentum permit a general constraint to be placed on 
the cumulative results of small perturbations by distant bodies over a long 
period of time, or of multiple close encounters, but this constraint is not 
powerful enough to be useful.  

It is, however, possible to draw useful conclusions about the results of a 
single close encounter between planets whose small (but nonzero) 
eccentricities and similar semi-major axes permit such an encounter.
I investigate the hypothesis that the pre-encounter orbits were nearly
circular, because that is expected if planets condense from a disc.  If
they were not nearly circular, then the question of the origin of their
observed eccentricity can only be deferred, not solved.

In \S2 I present the elementary constraints implied by the conservation laws.
\S3 discusses the validity of the two-body approximation for the encounter
between two planets orbiting a massive third body.  In \S4 I develop a
``parabolic approximation'' for such encounters, estimate the parameter
regime for which it is valid.  This approximation is used in \S5 to 
constrain the parameters of planetary systems which could produce the 
present observed eccentric orbits after such an encounter.  \S6 contains a 
brief summary discussion.
\bigskip
\centerline{2. Conservation Laws}
\medskip
The present observed planetary orbit has semi-major axis $a$ and eccentricity
$e$.  There is no evidence for a second planet (designated 2) in any of the
systems in which the observed planet has an eccentric orbit, so we assume
this planet has been lost.  The conservation of energy then permits 
constraints to be placed on the pre-encounter orbital parameters; 
no constraint can be obtained from the conservation of angular momentum
because the angular momentum of the escaped planet is indeterminate.

If the observed eccentric orbit was produced in
a single encounter, that encounter must have occurred at a distance from
the star between $a(1-e)$ and $a(1+e)$, the smallest and greatest distances
of the present orbit from the star.  The initial radius $r_1$ of the orbit
of the observed planet, assumed nearly circular, satisfies
$$a(1-e) \le r_1 \le a(1+e).  \eqno(1)$$
Energy must be removed from this orbit in order to expel planet 2.  The 
greatest amount of energy is released if $r_1 = a(1+e)$.  If the initial 
orbit of planet 2 had semi-major axis $a_2$, and it is presently unbound, 
then the condition that the energy to unbind it came from the reduction in 
semi-major axis of planet 1 (from $a(1+e)$ to $a$) is
$${m_2 \over m_1}{1 + e \over e} a \le a_2,\eqno(2)$$
where $m_1$ and $m_2$ are the masses of the two planets.

It is also necessary that the orbits intersect.  Assuming $r_1 = a(1+e)$
yields
$$a_2(1 - e_2) \le a(1+e) \le a_2(1 + e_2). \eqno(3)$$
Then
$$a{1 + e \over 1 + e_2} \le a_2 \le a{1 + e \over 1 - e_2}. \eqno(4)$$
The right hand inequality in (4) may be combined with (2) to yield
$$1 - e {m_1 \over m_2} \le e_2. \eqno(5)$$
Unfortunately, without knowledge of $m_1/m_2$ no quantitative bound can be
placed on $e_2$.  It is expected that $m_1/m_2 \ge 1$ (the lighter planet is
more likely to be lost), so that (5) may only be the trivial statement $e_2
\ge 0$.

It is possible to constrain $m_1/m_2$ by requiring that the impulse
transferred in the encounter be sufficient to scatter planet 1 from its 
initial circular orbit to the observed ellipse.  The exact statement of
this condition is algebraically unwieldy and depends on the unknown
parameters of the initial orbit of planet 2, but a simple (though not the
strictest) bound may be obtained by assuming planet 2 initially in a radial 
orbit with zero total energy.  Then the relative velocity of encounter
$$v = \left[{3GM \over a(1+e)}\right]^{1/2}, \eqno(6)$$
where $M$ is the stellar mass.  The momentum transfer depends on the 
scattering angle, but satisfies
$$\Delta p \le {2 m_1 m_2 v \over m_1 + m_2}. \eqno(7)$$
The momentum transfer required to scatter planet 1 into the observed orbit
$$\Delta p = \left[{GM \over a(1+e)}\right]^{1/2} m_1 \left[1-(1-e)^{1/2}
\right]. \eqno(8)$$
Combining these conditions yields
$${m_1 \over m_2} \le {2 \sqrt{3} \over 1-(1-e)^{1/2}} - 1. \eqno(9)$$
For $e = 0.63$ the result is $m_1/m_2 \le 7.8$, which is not strict enough
to give any useful information when substituted in (5).

In the preceding paragraph the pre-encounter eccentricity of the second
planet $e_2 \to 1$.  If this assumption is made the the encounter has not
solved the problem of the origin of eccentricity.  If, on the other hand,
both planets are initially in nearly circular (but slightly eccentric and 
intersecting) orbits, then $v$ is small, as must be $\Delta p$, failing to 
explain the large observed $e$.  This case will be dealt with in \S4.
\bigskip
\centerline{3. Two-Body Approximation}
\medskip
If three bodies interact with each other simultaneously, then no simple
results beyond the conservation laws apply.  However, if the interaction
of two of the bodies has a characteristic time short compared to the period
of their motion about the third, then the interaction of the first two may
be treated independent of the influence of the third.  This is the case for 
two planets, of mass much less than that of the star, which briefly pass 
close to each other in their orbits around the star, as well as for the more
familiar case of a short period satellite orbit or hierarchal triple star.

Considering a planetary encounter as a two-body scattering problem, define
its impact parameter $b$ and relative velocity at infinity $v$.  Then the
necessary condition for the two-body approximation is
$${b \over v} \ll \left({a^3 \over GM}\right)^{1/2}, \eqno(10)$$
where $a$ is now the distance of the encounter from the star (approximately
equal to the semi-major axes of the planetary orbits).  For planets in
orbits of slightly different semi-major axes or small but finite 
eccentricities
$$v \approx {1 \over 2}{\delta a \over a}\left({GM \over a}\right)^{1/2};
\eqno(11)$$
where $\delta a \equiv \max(\vert a_1 - a_2 \vert, e_+a, ia)$, $e_+
\equiv e_1 + e_2$ and $i$ is the relative 
inclination.  Henceforth I will assume the orbits to be coplanar and 
$\delta a \ll a$.  Then the necessary condition (10) becomes
$$b \ll \delta a. \eqno(12)$$
This condition can be met if the orbits intersect, as is possible if $e_+ a$
is greater than the difference in semi-major axes, and is implicitly assumed 
by the hypothesis that observed eccentric
orbits are produced by close encounters.  It is consistent with small
pre-encounter eccentricities, provided that the initial semi-major axes are
similar enough.  This is not observed for planets in the Solar System, but 
there is no problem of planetary eccentricity to explain there.  It is
found for small bodies such as asteroids.
\bigskip
\centerline{4. Parabolic Approximation}
\medskip
In two-body scattering problems it is frequently possible to make an
impulsive approximation, corresponding to small deflections, in which the
momentum transfer is computed as if the initial trajectories of the particles
continued undeflected through their interaction.  In the present problem the
opposite ``parabolic approximation'' can be made, in which the path of 
their separation vector is nearly parabolic, and their relative velocities
change sign as a result of their interaction.  The momentum transfer is
then independent of $b$ even if $b \to 0$ and the particles approach
arbitrarily closely and feel arbitrarily strong forces (in the planetary
problem it is necessary to exclude the extreme case of actual impact).  The
parabolic approximation is valid for impact parameters
$$b \ll {Gm \over v^2}, \eqno(13)$$
where $m = m_1+m_2$.

Planets in nearly circular orbits with semi-major axes differing by $\sim 
\delta a$ will have relative velocities
$$v \sim \left({GM \over a^3}\right)^{1/2}\delta a. \eqno(14)$$
Combining (13) and (14) and taking $b \sim \delta a$ yields a sufficient
condition for the validity of the parabolic approximation (it is, of course,
valid for smaller $b$):
$${\delta a \over a} \ll \left({m \over M}\right)^{1/3} \approx 0.1, 
\eqno(15)$$
where the numerical value is appropriate to Jupiter-like planets orbiting
Solar type stars.  The parabolic approximation is valid for all encounters 
(which may be defined as the overtaking of the longitude of one planet by 
that of the other) between planets whose semi-major axes satisfy (15).  It
is also valid for encounters with $b \ll m a^3/M \delta a^2)$ between 
planets with larger $\delta a$ (this may be considered the definition of 
intersecting orbits).

Equation (15) must hold if there are to be close encounters between planets
in nearly circular ($e_+ < (m/M)^{1/3} \approx 0.1$) orbits.  Hence the 
parabolic approximation is always applicable to the problem discussed in 
this paper.  If it is not valid then at least one of the initial orbits was
significantly eccentric, and the encounter has not produced an eccentric 
orbit from nearly circular orbits.  
\bigskip
\centerline{5. Results}
\medskip
The parabolic approximation simplifies the dynamics, because in a parabolic 
encounter the planetary velocities (in the frame of their center of mass) 
simply reverse sign.  It is then possible to relate the observed eccentricity
to the pre-encounter eccentricity $e_2$ of planet 2, assuming planet 1 was 
initially in an exactly circular orbit of radius $a(1+e)$.  The
assumption of small eccentricities implies that the pre-encounter velocity
of planet 2 was nearly circumferential.  The requirement that planet 2 be
expelled is most easily met if $a(1+e)=a_2(1-e_2)$, in which case both
velocities are exactly circumferential.  The required momentum transfer is
given by Equation (8).  The relative velocity is then
$$v = {m_1 +m_2 \over 2 m_2} \left[{GM \over a(1+e)}\right]^{1/2} u, 
\eqno(16)$$
where $u \equiv 1 - (1-e)^{1/2}$.  Using the pre-encounter circular 
Keplerian velocity of planet 1, the pre-encounter velocity of planet 2 may 
be calculated, and from that its total energy:
$$E_2 = {GM m_2 \over a(1+e)} \left(-{1 \over 2} - {u \over 2
\mu} + {u^2 \over 8 \mu^2}\right), \eqno(17)$$
where the mass fraction $\mu \equiv m_2/(m_1+m_2) < 1$.  Similarly, the 
pre-encounter angular momentum of planet 2 is
$$L_2 = \big[GMa(1+e)\big]^{1/2} m_2 \left(1 - {u \over 2 \mu} \right). 
\eqno(18)$$
From these quantities its eccentricity is obtained (Symon 1960):
$$e_2 = \big(4 u^2 -4 u^3 + u^4\big)^{1/2}. \eqno(19)$$
In the limit $e \to 0$ 
$$e_2 \to {e \over 2 \mu}. \eqno(20)$$

In order to evaluate Equations (19) or (20) explicitly it is necessary to
know $\mu$.  For $m_1 = m_2$ ($\mu = 1/2$) Equation (19) reduces to $e_2 =
e$; the planets simply exchange orbits.  This neither solves the eccentricity
problem nor is consistent with the expulsion of planet 2.  If $m_1 > m_2$
then $e_2 > e$ so that a larger eccentricity is required before the 
encounter than is found after it.  If $m_1 < m_2$ then $e_2 > e/2$; the 
pre-encounter eccentricity may be smaller than that observed, but by no more
than a factor of two. 

The condition (5) on $m_1/m_2$ must be satisfied in order than planet 2 be
lost.  From this condition it is possible to calculate (iteratively) a lower
bound on $e_2$ as a function of $e$.  The results are shown in Figure 1.
For $e < 0.5$ the minimum $e_2$ ($e_2 > e$) is found for $m_1 > m_2$.  For
$e > 0.5$ solutions are possible with $e_2 < e$; these solutions have $m_1
< m_2$, but large $e_2$.  If $m_1 \ge m_2$ is required then $e_2 \ge e$.  In
computing Figure 1 the parabolic assumption was made, which is not valid for
all encounters permitted when the eccentricity is large.  However, 
non-parabolic encounters transfer less momentum than parabolic ones (the
scattering angle is less than $180^\circ$), and therefore require greater
relative velocity and larger pre-encounter eccentricity than the assumed
parabolic encounters.
\bigskip
\centerline{6. Discussion}
\medskip
It is not possible to produce the observed eccentric orbits by single 
encounters between planets initially in nearly circular orbits.  This 
conclusion is almost obvious: scattering from a circular orbit to one with 
substantial eccentricity requires substantial momentum transfer, and hence a
substantial relative velocity between the two planets, which cannot be 
obtained if they are are initially in coplanar and nearly circular orbits of
nearly the same size.

How were the observed eccentricities produced?  It is not possible to make
any general arguments concerning the effects of small perturbations acting
over the age of the planetary system, or constraining the effects of possible 
repeated close encounters.  Mazeh and Krymolowski (1996) have suggested that
the eccentricity of the planet orbiting 16 Cyg B may be attributable to the 
presence of the distant companion star 16 Cyg A in an (assumed) inclined
orbit, while Rasio and Ford (1996) have suggested that single planets in 
tightly bound orbits could result from the long term interaction of planets
in orbits initially satisfying Eq. (15).  These numerical results illustrate
the potential richness of orbital evolution produced by the integration of 
small forces over long times.  Certainly (and fortunately) the major planets
do not induce such large eccentricities in Earth's orbit, or their own.

The possibility that a mechanical collision between two planets could 
produce an eccentric orbit may also be dismissed.  Such a collision between 
Jupiter-like gaseous planets would be dissipative, and would have the effect
of averaging their orbital parameters.  If their initial orbits were nearly 
coplanar and nearly circular the result could only be a nearly circular orbit.

It may be useful to consider the early history of the eccentric planetary
systems.  For example, tidal interaction with a rapidly spinning star might
make the eccentricity of a planetary orbit grow, as the accelerating tidal
torque is largest at periastron.  Tidal interaction is negligible when the 
star is of Solar dimensions and $a$ is an AU or more, but may have been 
significant during the pre-main sequence or protostellar stages of 
evolution, when the star was larger.  Eccentricity induced by a rapidly 
rotating star has a characteristic signature: close planets have larger 
eccentricities than more distant ones.  The applicability of this process to
systems like 16 Cyg B would be tested if additional more distant planets are
found accompanying the observed eccentric planets, and it is consistent with
the absence of planets in orbits inside the observed eccentric orbits.  It
might explain the comparatively large ($e = 0.206$) eccentricity of Mercury.  

I thank T. Mazeh and T. Piran for discussions, NSF AST 94-16904 and NASA
NAG 52682 for support, the Hebrew University for hospitality and a
Forchheimer Fellowship, and Washington University for the grant of
sabbatical leave.
\vfil
\eject
\centerline{References}
\parindent=0pt
\def\ref{\medskip \hangindent=20pt \hangafter=1}
\ref
Cochran, W. D., Hatzes, A. P., Butler, R. P. \& Marcy, G. W. 1996, preprint
astro-ph/9611230
\ref
Marcy, G. W. \& Butler, R. P. 1996, ApJ 464, L147
\ref
Mazeh, T., \& Krymolowski, Y. 1996, preprint
\ref
Mazeh, T., Latham, D. W. \& Stefanik, R. P. 1996, ApJ 466, 415
\ref
Rasio, F. A., \& Ford, E. B. 1996, Science 274, 954
\ref
Symon, K. R. 1960 {\it Mechanics} (Addison-Wesley, Reading, Mass.)
\vfil
\eject
\centerline{Figure Caption}
\medskip
Figure 1: Allowed initial eccentricity $e_2$ of unseen planet 2 as a function
of observed eccentricity $e$.  The planetary mass ratio $m_1/m_2$ has been
found implicitly, assuming that planet 2 was expelled as a result of
the encounter.  The parabolic approximation was made, and planet 1 was
assumed initially in a circular orbit of radius $a(1+e)$.
\vfil
\eject
\bye
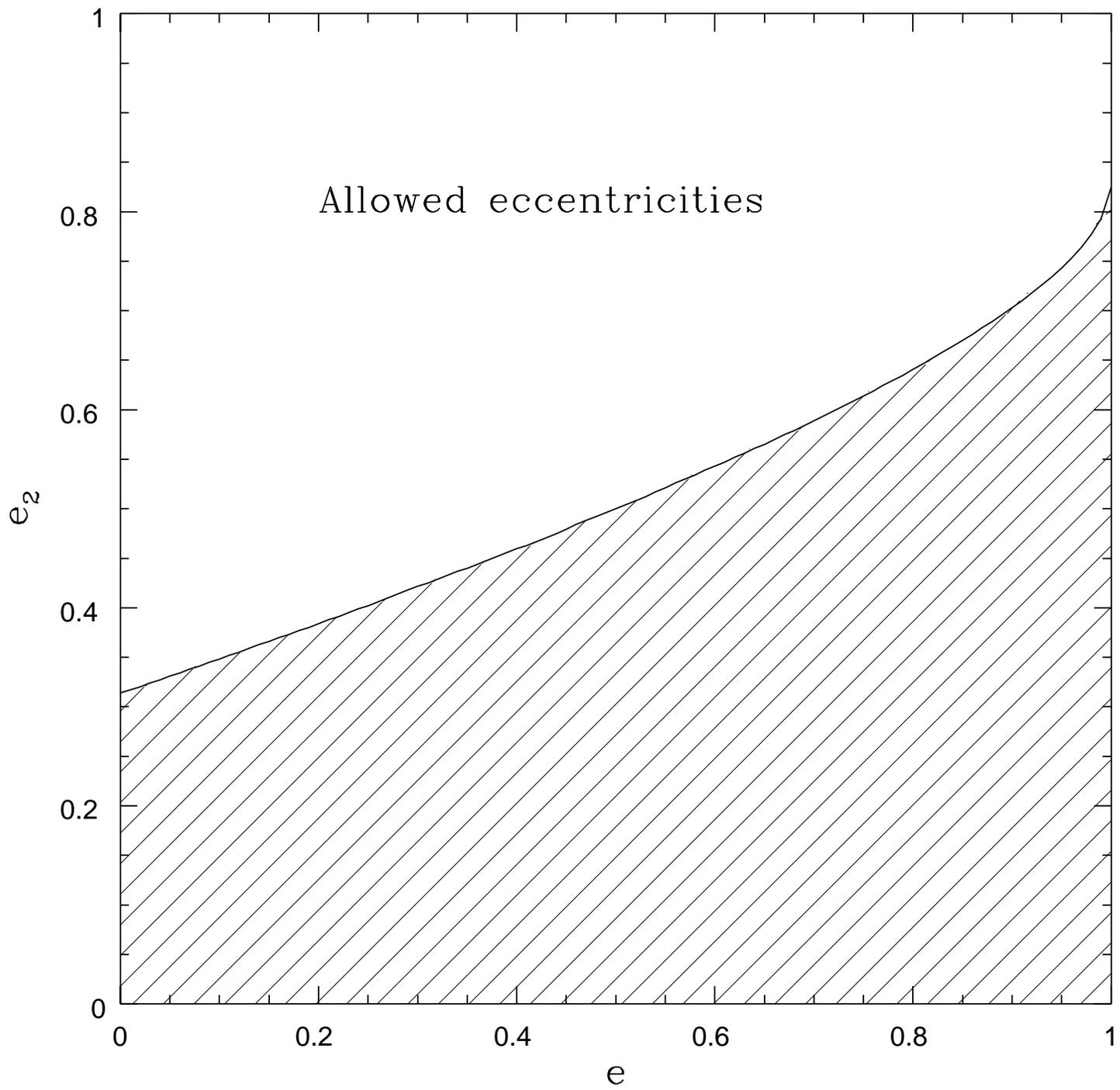
\end